\documentclass[prd, twocolumn,preprintnumbers, groupedaddress]{revtex4-1}
\usepackage{graphicx}
\usepackage[mathlines]{lineno}
\usepackage{amsmath}
\usepackage{ulem}

\usepackage{slashed}
\usepackage{natbib}
\usepackage{bm}
\usepackage{amsfonts}
\usepackage{amssymb}
\usepackage{booktabs}
\usepackage{epsfig}
\usepackage{subfigure}
\usepackage{hyperref}
\usepackage{soul}
\hypersetup{
	colorlinks=true,
	citecolor=blue,
	linkcolor=blue,
	urlcolor=blue}
\everymath{\displaystyle}
\usepackage[usenames,dvipsnames]{color}
\usepackage{bm}
\definecolor{CB}{rgb}{0.2, 0.3,1}

\newcommand{\bea}{\begin{eqnarray}}
\newcommand{\eea}{\end{eqnarray}}

\begin{document}
	\title{Effect of background magnetic field on the normal modes of conformal dissipative chiral hydro and a novel mechanism for explaining pulsar kicks}
	\author{Manu George}
	\email[e-mail: ]{kuttan.mgc@gmail.com}
	\author{Arun Kumar Pandey}
	\email[e-mail: ]{ arunp77@gmail.com}
	\affiliation{%
		\centerline{Physical Research Laboratory, Theory Division, Ahmedabad 380009, Gujarat, India}\\
	}
	\bigskip
\begin{abstract}
	In the present work, we have studied the collective behaviour of a chiral plasma with first and second order dissipative corrections allowed by conformal symmetry. We have derived dispersion relations for ideal, first order and second order   MHD for the chiral plasma. We have also used our results to explain the observed pulsar kicks.
 \end{abstract}
 \maketitle
\section{Introduction}
Relativistic hydrodynamics is useful in understanding several  physical systems such as early Universe, astrophysical systems, quark gluon plasma etc.
%
The evolution equations for the variables of the theory viz. local velocity $u^{\mu}(x)$, temperature $T(x)$ and chemical potential $\mu(x)$ associated with the conserved charges are derivable from conservation of the energy-momentum tensor. It is well known from quantum field theories that the conservation of chiral current is spoiled by parity violating quantum effect known as the chiral anomaly \cite{vilenkin:1969, bell:1969}. Recently, kinetic and hydrodynamic theories have been modified to incorporate the parity odd effect appropriately \cite{Son:2012zy, Son:2009tf}. Example for a plausible system that can be studied using chiral hydrodynamics is the plasma in the early Universe, when there is asymmetry in the left and right handed particle number densities. In the presence of either an external gauge field or rotational flow of the fluid, the parity may be broken and there may be a current in the direction parallel to external magnetic field or in the direction parallel to the vorticity generated due to rotational flow in the chiral plasma. These two currents are known as ``Chiral Magnetic Current" (CMC) and ``Chiral Vortical Current" (CVC) respectively. The effect due to which CMC and CVC appears, are called as ``chiral magnetic effect (CME)" \cite{Tashiro:2012mf,Basar:2012gm, Fukushima:2012vr} and ``chiral vortical effect(CVE)" \cite{Vilenkin:1979ui} respectively. The transport coefficients associated with these currents can be parametrized in terms of chiral chemical potential $\mu_A=(\mu_R-\mu_L)$. These  effects 
vanish in the absence of any net chiral imbalance within the system. These transport coefficients are calculated by several authors by imposing second law of thermodynamics $\partial_{\mu} s^{\mu} \geq  0$, where $s^{\mu}$ is the entropy four current \cite{Son:2009tf, Erdmenger:2009, Banerjee:2011}. There are many attempt to explain the chiral effects using modified kinetic theory \cite{Son:2012dy, Stephanov:2012my, chen:2013jp, Buividovich:2009wi, Buividovich:2010tn} and it has been shown that there exist collective excitations (different from usual density waves in the  standard plasma), called chiral magnetic waves (CMW) \cite{Kharzeev:2011ey, Newman:2005hd} and chiral vortical wave (CVW) \cite{Jiang:2015cva}. Also, there can  be gap-less excitations in the presence of dynamical electromagnetic (colour) fields called chiral plasma instability (CPI) \cite{Sadofyev:2011, Akamatsu:2014uu,  Joyce:2012mm, Boyarsky:2012fr, Yamamoto:2015ns}. In a recent work \cite{Yamamoto:2015ns}, author has found a new type of gap-less collective excitation induced by chiral effects in an external magnetic field. This is a transverse wave known as chiral Alfv\'en wave and it exists even in the incompressible fluids. It  is shown that, these transverse modes, get split when we take account the effect of first order viscous terms along with the ideal terms in the energy momentum tensor \cite{Abbasi:2015saa}. However a detailed study of first and second order conformal chiral viscous hydrodynamics has not been done. In the present work, we have derived dispersion relations using the first and second order conformal viscous hydrodynamics theory in presence of the external background magnetic fields for the chiral plasma.

At this juncture its worth introducing  the pulsar kicks, as later we use our results from normal mode analysis of chiral magnetohydrodynamics to put forward a novel mechanism to explain it. It was observed that neutron stars often does not move with the same velocity of its progenitor star \cite{lyne1982proper,gott1970runaway,iben1996origin,hansen1997pulsar}, but rather with a substantially greater speed. There are several attempts to explain these observations. Interested readers are directed to the Refs. \cite{iben1996origin,kusenko1999pulsar,kusenko2004pulsar,Kaminski:2014jda}. The exact reasons for the pulsar kick is  not known. However, it is believed that it must be because of the way in which supernovae explode. A proper understanding of the cause for pulsar kick may give more insight into the supernova explosion mechanisms. A particular interest in this  direction of investigation is to understand whether there is some correlation between the direction of  kick velocity and angular momentum, alignment of the magnetic field etc. Though there is no clear idea about these topics yet, it is worth investigating for we believe that, later observations may shed more light onto  these questions.

In our discussion, we have adopted $(-,+,+,+)$ signature for the metric, generally used in cosmology and the conformal metric is given by
\begin{equation}
	ds^2= g^*_{\mu\nu}dx^\mu d^\nu= a(\tau)^2(-d\tau^2 + g^{ij} \,dx_i d_j), \label{eq:FLRW-met}
\end{equation}
where $\tau$ denotes the conformal time and comoving proper $t$ time is given in terms of the conformal time as $\tau=\int dt/a(t)$. One of the important parameter is Hubble parameter  $H(t)=\dot{a}/a$ (here $\dot{a}=da/dt$), which gives the rate of expansion of Universe. The metric given in equation (\ref{eq:FLRW-met}) is also known as conformally flat metric as it can be written in terms of flat space Minkowski metric $\eta_{\mu\nu}$ as $g^*_{\mu\nu}= \eta_{\mu\nu}=a^{-2}\,g_{\mu\nu}$. It has been shown that, under a conformal transformation, evolution equations of the fluid remain invariant and one can transform the evolution equations to a flat space evolution equation by redefining the variables, for example magnetic fields and electric fields as $a^{-2} {\bm B}$ and $a^{-2} {\bf E}$ respectively. Similarly, other thermodynamic variables are scaled with scale factor in expanding conformal flat space-time as: $\sigma\rightarrow a^{-1}\sigma$, $\mu\rightarrow a^{-1}\mu$ and $T\rightarrow a^{-1}T$ \cite{MacDonald:1982, Detmann:1993}. With the new variables, Maxwell's equation remain same in the conformally flat space-time. In the following discussions,  we will work with the above defined comoving quantities.

The rest of this paper is organized as follows: in section (\ref{sec-chMHD}), we have discussed conformal fluid hydrodynamics for the chiral plasma using a ideal and viscous hydrodynamics. In  section (\ref{sec-Dis}), dispersion relation for the transverse and longitudinal modes in the case of ideal as well as viscous Hydrodynamics has been derived. In section (\ref{sec-pulsarkick}), we put forward a novel mechanism to explain the pulsar kick using results obtained in the previous section. Last section, is dedicated for discussion and results of the present work.
\section{Chiral Magnetohydrodynamics} \label{sec-chMHD}
The dynamics of the chiral fluids can be obtained by following covariant evolution equations:
\begin{gather}
	\nabla_{\mu} T^{\mu\nu}= F^\nu_\lambda j^{\lambda}_V \label{conservT}\\
	\nabla_{\mu}j^{\mu}_{L,R}=C_{L,R}E^{\mu}B_{\mu}, \label{conservJ}
\end{gather}
where $T^{\mu\nu}$, $j_V^{\mu}$ and $j_{L,R}^\mu$ are the energy momentum tensor and 4-vector and left/right handed currents respectively. The symbol $\nabla_\mu$ is covariant derivative and is defined as $\nabla_\nu A^\mu =\partial_\nu A^\mu+ \Gamma^\mu_{\nu\sigma}\, A^\sigma$. The left- and right- handed currents are given by
\begin{gather}
j^{\mu}_{L,R}= n_{L,R}u^{\mu}+\nu^{\mu}_{L,R}, \label{eq:LR-current}
\end{gather}
where $n_{L,R}$ is the charge chiral charge density of the left handed and right handed particles in the fluid. However,  $\nu^{\mu}_{L,R}$ is given by
\begin{equation}
\nu^\mu_{L,R}=\frac{1}{2}\sigma E^{\mu} +\xi_{L,R}\omega^{\mu}+\xi_{L,R}^{B}\, B^\mu \label{eq:nuRL}
\end{equation}
In the present work, we have considered a 4-dimensional space-time. Greek indices run from $0$ to $4$. However, the Latin indices are used for the 3-D spatial-coordinates  and are given by ($1, 2, 3$). In this 4-D space-time, the electric $E^\mu$ and magnetic $B^\mu$ four vectors are totally spatial vectors and they are defined in terms of electromagnetic field strength tensor  $F^{\mu\nu}=\nabla_{\mu} A_{\nu}-\nabla_{\nu} A_{\mu}$ and 4-velocity of the fluid $u^\mu$ as: $E^{\mu}=F^{\mu\nu}u_\nu$, $B^{\mu}=\frac{1}{2}\epsilon^{\mu\nu\alpha\beta}\, u_\nu F_{\alpha\beta}$. Here $A^\mu$ being the electromagnetic 4-potential and $u^\mu$ satisfy $u_\mu u^\mu=-1$. Maxwell's equation can be obtained by generalized Maxwell's equations:
\begin{equation}
\nabla_\nu F^{\mu\nu}=j^{\mu}_V~~~~ \text{and} ~~~~~ F_{[\mu\nu,\lambda]}=0 \label{eq:Max}
\end{equation}
One of the novel feature of the chiral fluid is that, it carries both electric currents $j_V^{\mu}=\bar \psi \gamma^{\mu}\psi$ and chiral/axial currents 
$j^{\mu}_A=\bar \psi \gamma^{\mu}\gamma^{5}\psi$ (here $\psi$ denotes the fermionic fields). In the basis of the vector and axial currents, the left- and right-handed currents can be written as: $j^{\mu}_R=\frac{1}{2}(j^{\mu}_V+j^{\mu}_A)$ and $j^{\mu}_L=\frac{1}{2}(j^{\mu}_V-j^{\mu}_A)$.
Therefore, one can write the Vector and axial current using the equations (\ref{eq:LR-current})-(\ref{eq:nuRL}) as
\begin{eqnarray}
j_V^{\mu} & = &n_V u^\mu+ \sigma E^{\mu}+\xi\omega^{\mu}+\xi^{B}B^\mu, \label{eq:currentV} \\
j_A^{\mu} & = & n_A u^\mu+\xi_A^{\prime}\omega^{\mu}+\xi_A^{\prime B}B^\mu , \label{eq:currentA}
\end{eqnarray}
here the transport coefficients $n_V$, $n_A$, $\xi$, $\xi^B$, $\xi_A$ and $\xi_A^{B}$ are give by $n_V=n_R+n_L$, $n_A=n_R-n_L$,
$\xi=\xi_R+\xi_L$, $\xi^B=\xi_R^B+\xi_L^B$, and $\xi_A=\xi_R-\xi_L$, 
$ \xi_A^{B}=\xi^{B}_R-\xi^{ B}_L$. 
The first two terms in $j_V^\mu$ are conventional drift and OHM current respectively in the electrodynamics, which describes, how electric current flow in the direction of electric field. 
$\xi_{R,L}$ and $\xi_{R,L}^B$ are transport coefficients corresponding to CME and CVE and they are function of chemical potentials of the right- and left-handed particles of the chiral particles and the temperature. 
%
\subsection*{Chiral viscous fluid}
For a viscous fluid, energy momentum tensor $T^{\mu\nu}$ is defined as $T^{\mu\nu}=T^{\mu\nu}_{\rm ideal}+\tau^{\mu\nu}_{(1)}+\tau^{\mu\nu}_{(2)}$. Here $T^{\mu\nu}_{\rm ideal}$ represents ideal fluid contribution to the total energy momentum tensor. However, two other terms $\tau^{\mu\nu}_{(1)}$ and $\tau^{\mu\nu}_{(2)}$ are the first and second order contributions. The viscous terms must follow $\tau^{\mu\nu}_{(1)} \, u_\nu=\tau^{\mu\nu}_{(2)}\, u_\nu =0$. The ideal part of the energy momentum tensor is given by
\begin{equation}
T^{\mu\nu}_{\rm ideal}=(\varepsilon+p)u^\mu u^\nu+p g^{\mu\nu}, \label{T-munu}
\end{equation}
The most general form of the first order and second order viscous terms, which follows the conformal transformations are given is given as \cite{Kharzeev:2011kh, Erdmenger:2009, Banerjee:2011}
\begin{eqnarray}
\tau^{\mu\nu}_{(1)} & = & -2\eta \, \sigma^{\mu\nu}
\label{eq:Istorder}\\
\tau_{(2)}^{\mu\nu} & = & \lambda_1 ~ \Pi^{\mu\nu}_{\alpha\beta}\, \nabla^\alpha \omega^\beta +\lambda_2~
\Pi^{\mu\nu}_{\alpha\beta}\, \omega^\alpha \nabla^\beta \bar{\mu} \nonumber \\
& + & \lambda_3~\Pi^{\mu\nu}_{\alpha\beta}\, 
\epsilon^{\gamma\delta\eta\alpha}\, \sigma^\beta_\eta\, u_\gamma \nabla_\delta \bar{\mu} + \lambda_4\, \Pi^{\mu\nu}_{\alpha\beta} \nabla^\alpha B^\beta\nonumber \\
& + & \lambda_5 ~\Pi^{\mu\nu}_{\alpha\beta}\, B^\alpha \nabla^\beta \bar{\mu} + \lambda_6~ \Pi^{\mu\nu}_{\alpha\beta}\,  E^\alpha \, B^\alpha \nonumber\\
& + & \lambda_7 ~ \Pi^{\mu\nu}_{\alpha\beta} \, \epsilon^{\gamma\delta\eta\alpha} \, \sigma^{\beta}_\eta u_\gamma\, E_\delta + \lambda_8~
\Pi^{\mu\nu}_{\alpha\beta}\,  \omega^{\alpha}\, E^{\beta}
\label{eq:IIndorder}
\end{eqnarray}
Here, $\sigma_{\mu\nu}$ and the vorticity four vectors are related by the relation $\nabla_{\mu} u_\nu= \sigma_{\mu\nu}+\omega_{\mu\nu}$ with $\omega_{\mu\nu}=\frac{1}{2}(\nabla_{\mu} u_\nu- \nabla_{\nu} u_\mu)$. The coefficients $\eta$ and $ \lambda_1,  \lambda_2 ....,  \lambda_8$ in the expressions of energy momentum tensor are transport coefficients and the projection operator $\Pi^{\mu\nu}_{\alpha\beta}$ is defined in terms of spatial projection operators $P_{\mu\nu}$ as $\Pi^{\mu\nu}_{\alpha\beta}=\frac{1}{2}\large[P_{\alpha}^{\mu} P_{\beta}^{\nu}+P_{\beta}^{\mu} P_{\alpha}^{\nu}-\frac{2}{3}P^{\mu\nu} P_{\alpha\beta}\large]$. $\bar \mu$ is represents the ratio of the chiral chemical potential and the temperature ($\mu_A/T$). Levi-Civita tensor $\epsilon^{\mu\nu\alpha\beta}=1/2 \epsilon^{\mu\nu\alpha\beta}u_\nu \nabla_\alpha u_\beta$,  is a purely antisymmetric tensor. In the present work, we have considered the second order viscous corrections, arising from the triangle anomaly \cite{Kharzeev:2011kh}, which in turn leads to the modification of the standard electrodynamic magnetic modes in the fluid.
\section{Dispersion relations} \label{sec-Dis}
In the upcoming subsections, we have derived dispersion relations for the ideal and viscous chiral fluid using the equations (\ref{conservT}) and (\ref{conservJ}). The evolution equations are
\begin{eqnarray}
	u^\mu \partial_{\mu}\varepsilon &  = & j^\tau\, u_\nu\,F^\nu_\tau \label{eq:energycon}\\
   (\varepsilon+p)\, u^\mu\,\partial_\mu u^\nu & = & \partial^\mu p + P^\nu_\mu F^{\mu\rho} j_\rho + \partial_{\mu} \tau^{\mu\nu} \label{eq:velcon},
\end{eqnarray}
here $\tau^{\mu\nu}=\tau^{\mu\nu}_{(1)}+\tau^{\mu\nu}_{(2)}$.
We have derived the dispersion relations for a ideal chiral fluid and viscous fluids in the next two subsections.
\subsection{Splitting of chiral magnetic modes of a ideal chiral plasma in presence of background magnetic field}
In the present subsection, we have revisited the dynamics of the chiral plasma and we have derived the dispersion relations for the ideal chiral fluid. We have rederived evolution equation of the chiral fluid using (\ref{conservT}) and (\ref{conservJ}).
We consider the following counting schemes: $\partial_t\sim O(\epsilon_t)$, $\bm v\sim O(\delta)$ and  $\bm\nabla,~\bm B\sim O(\epsilon_s)$.  From equations (\ref{eq:energycon}-\ref{eq:velcon}), one can have the following expressions, in absence of the viscous terms \cite{Yamamoto:2015ns};
\begin{eqnarray}
\left(\frac{\partial}{\partial \tau}+\bm v\cdot\bm{\nabla}\right)\varepsilon+(\varepsilon+p)\bm{\nabla\cdot\bm v}& = & 0\, \label{energyconsunpert}\\
(\varepsilon+p)\left(\frac{\partial}{\partial \tau}+\bm v\cdot\bm{\nabla}\right)\bm v & = & - \bm{\nabla}p +\bm j\times\bm B \, \nonumber \\
- \bm v(\frac{\partial}{\partial \tau}
+\bm v\cdot\bm{\nabla})p
\label{momentumconsunpert}
\end{eqnarray}
together with the constituent equation,
\begin{equation}
\frac{\partial n}{\partial \tau}+\bm\nabla\cdot\bm j=0,
\end{equation}
where,
\begin{equation}
\bm j=n\bm v+\xi\bm\omega+\xi_B\bm B. \label{eq:Currenttotal}.
\end{equation}
 Here $n$ is the  chiral charge density. In the conformally flat FLRW metric, we can obtain the Maxwell's equation using equation (\ref{eq:Max}) as
\begin{gather}
\frac{\partial {\boldsymbol{B}}}{\partial \tau}+\boldsymbol{\nabla}\times\boldsymbol{E}=0 \, , \label{eq:Max1}\\
{\boldsymbol{{\nabla}\cdot E}}=4\pi n \,,
\label{eq:Max2} \\
{\boldsymbol{{\nabla}\cdot{B}}}=0 \, ,
\label{eq:Max3}\\
{\boldsymbol{{\nabla}\times{B}}}= 4\pi {\boldsymbol{J}}+\frac{\partial {{\boldsymbol{E}}}}{\partial \tau} \, ,
\label{eq:Max4}
\end{gather}
here $n$ is the total charge density. Above ${\bm B}$, ${\bm E}$, $n$ and ${\bm J}$ are all quantities in the conformally flat space. In order to  do the  linear analysis, we take:
$\varepsilon =\varepsilon_0 +\delta \varepsilon$, $p=p_0 +\delta p $, $ n=n_0+\delta n$, $u^\mu =u_0^\mu+\delta u^\mu = (1,\bm 0)+(0, \delta \textbf{v})$ (here we have taken Lorentz factor $\gamma \approx 1$), 
$T^{\mu\nu} =T^{\mu\nu}_0 +\delta T^{\mu\nu}$, $\bm E=\delta\bm E$ and $\bm B=\bm B_0+\delta\bm B$. 
Subscript ``0" denotes the background values of the corresponding quantities. Substituting these back to the (\ref{energyconsunpert}) and (\ref{momentumconsunpert}) and keeping only the terms linear in perturbations, we obtain,
\begin{align}
&\frac{\partial\, \delta \varepsilon }{\partial\, \tau}+(\varepsilon_0+p_0)\nabla\cdot\delta\bm v=0\label{energyconspert}\\
&
(\varepsilon_0+p_0)\frac{\partial\, \delta\bm v}{\partial \tau}+v_s^2\bm\nabla\,\delta \varepsilon +\bm B_0\times\big(\bm\nabla\times\delta\bm B-\xi_0\,\bm{\omega}\big)=0
\label{momentumconspert}
\end{align}
where, $v_s=(dp/d \varepsilon)^{1/2}$ is the speed of sound in the medium and $\bm \omega=\bm\nabla\times\bm v$ is the vorticity. The above expressions  have to be supplemented  with Maxwell's equations. In the infinite conducting (ideal plasma) limit  Maxwell's equations lead to \cite{Pandey:2017zpg},
\begin{equation}
\frac{\partial\delta\bm B}{\partial \tau}=\bm{\nabla}\times\big(\delta\bm v\times\bm B_0\big)
\label{Maxwpert}
\end{equation}
Differentiating (\ref{momentumconspert}) with respect to time and using (\ref{Maxwpert}) we get,
\begin{align} 
\frac{\partial^2\delta\bm v}{\partial \tau^2} & -v_s^2\bm \nabla(\bm{\nabla}\cdot\delta\bm v)+\bm v_A\times\big[\bm \nabla\times(\bm\nabla\times(\delta\bm v\times \bm v_A))\big]- \nonumber \\ & \frac{\xi_0}{(\varepsilon+p_0)^{1/2}}(\bm v_A\times\bm \nabla)\times\frac{\partial\delta\bm v}{\partial \tau}=0
\label{mastereqnconfig}
\end{align}
where, $v_A=B_0/(\varepsilon_0+p_0)^{1/2}$ is  speed of the Alfv\'en wave. To study the normal modes, we take  $\delta\bm v(t, \bm x)=\bm \delta v_{\omega,{\bf k}}\, e^{-i(\omega \tau-{\bf k}\cdot{\bf x})}$. Thus  Eq. (\ref{mastereqnconfig})  take following form in  the Fourier space.
\begin{align}
-\omega^2\, & \delta v_{\omega,{\bf k}}+ (v_s^2+v_A^2)({\bf k}\cdot \bm \delta v_{\omega,{\bf k}}){\bf k}+({\bf k}\cdot \bm v_A) \Big[({\bf k}\cdot\bm v_A)\, 
 \delta  v_{\omega,{\bf k}}\nonumber \\
&-(\bm v_A\cdot\bm \delta v_{\omega,{\bf k}}){\bf k} 
- ({\bf k} \cdot \bm \delta v_{\omega,{\bf k}})\bm v_A\Big]-\frac{\xi_0\, \omega}{(\varepsilon_0+p_0)^{1/2}}
\nonumber \\
&\times\Big[(\bm v_A\cdot\bm \delta v_{\omega,{\bf k}}) {\bf k} - (\bm v_A\cdot{\bf k})\bm \delta 
v_{\omega,{\bf k}}\Big]= 0
\end{align}
When the propagation vector $\bm k$ is perpendicular to the background magnetic field $\bm B_0~(\bm v_A)$, which means that $\bm v_{\omega, {\bf k}}\parallel {\bf k}$, we have from the above expression,
\begin{equation}
	\omega=\pm\sqrt{v_s^2+v_A^2}~k
\end{equation}
which corresponds to the mixing of Alfv\'en modes with sound waves present in occur standard plasma generally known as the magneto-sonic waves. For a transverse perturbation travelling along the background field, we get,
\begin{equation}
	\omega=-\frac{\xi v_A}{2(\varepsilon_0+p_0)^{1/2}}k\pm k\sqrt{\frac{\xi^2 v_A^2}{4(\varepsilon_0+p_0)}+v_A^2}
	\label{chiralalfvsplitmode}
\end{equation} 
In the absence of any chiral vorticity ($\xi=0$) this mode reduces to standard Alfv\'en mode $\omega=\pm v_A k$, corresponds to the propagation of wave along and opposite to $\bm B_0$. Thus from Eq.(\ref{chiralalfvsplitmode}), we observe that the standard Alfv\'en mode split in the chiral plasma. Furthermore, both the modes travel with different group velocity viz. $v_g^{\mp}=\Big|\frac{\xi v_A}{2(\varepsilon_0+p_0)^{1/2}}\mp \sqrt{\frac{\xi^2 v_A^2}{4(\varepsilon_0+p_0)}+v_A^2}\Big|$. In the upcoming section, we will discuss the importance of the splitting of the wave modes in details.
\subsection{Chiral Hydrodynamics with first order hydrodynamics of a chiral plasma}
In the present section, we have done linearization of equation (\ref{conservT}) in presence of background magnetic fields using first order dissipative viscous corrections for the chiral plasma.  Using equation (\ref{conservT}), the evolution equation with viscous effect can be written as:
\begin{gather}
\frac{\partial \delta \varepsilon}{\partial \tau} +(\varepsilon_0+p_0)\nabla\cdot \left(\delta\textbf{v}\right)=0 \label{eq:Istvisepsilon}\\
(\varepsilon_0+p_0)\frac{\partial\, \delta\bm v}{\partial \tau}+v_s^2\bm\nabla\,\delta  \varepsilon +\bm B_0\times\big(\bm\nabla\times\delta\bm B-\xi_0\,\bm{\omega}\big) \nonumber \\
 = -\eta_0 [\nabla^2 \delta\bm v -\nabla (\nabla\cdot \delta\bm v)] -\frac{1}{3}\eta_0\nabla \left(\nabla\cdot \delta\bm v\right). \label{eq:IstvisV}
\end{gather}
 Similar to the previous section, we can obtain dispersion relation for the transverse and longitudinal propagation.
 To do so, we have taken Fourier transform of the time derivative of the equation (\ref{eq:IstvisV}) and considered following two cases: $\textbf{(i).}$  ${\bf k}\perp \bm B_0$ and $\textbf{(ii).}$  ${\bf k} \parallel \bm B_0$.

\textbf{Case~i.}   ${\bf k}\perp \bm B_0$,
 \begin{equation}
 \omega =\frac{i\, \eta_0 k^2}{3(\varepsilon_0+p_0)}\pm \frac{1}{2} \sqrt{\left(\frac{2\, i\, \eta_0 k^2}{3(\varepsilon_0+p_0)}\right)^2 + 4 (v_s^2+ v_A^2) \, k^2} \label{eq:fir-or-vis}.
 \end{equation}
 Above we have used time derivative of $\delta\varepsilon$ from equation (\ref{eq:Istvisepsilon}) in equation (\ref{eq:IstvisV}). Second term represents the first order viscous effects and modes will damp exponentially due to this term. However, third term proportional represents the mixing of the Alf\'ven and sound modes. This mixed mode will propagates with the $\sqrt{(v^2_s+v_A^2)}$. However, due to the first order viscous effect in the chiral fluid, modes will damp exponentially.
 
\textbf{Case~ii.}  when ${\bf k}\, \parallel \, \bm B_0$\\ 
In this case, we consider $\delta \bm v_{\omega, {\bf k}}\, \parallel\,\bm B_0$ and $\delta \bm v_{\omega, {\bf k}} \perp \bm B_0$. 

$\textbf{ii(a)}.$ For (${\bf k}\, \parallel \, \bm B_0$ and $\delta \bm v_{\omega, {\bf k}}\, \parallel\,\bm B_0$):
\begin{eqnarray}
\omega = \frac{i\, \eta_0 k^2}{3(\varepsilon_0+p_0)}\pm \frac{1}{2} \sqrt{\left(\frac{2\, i\, \eta_0 k^2}{3(\varepsilon_0+p_0)}\right)^2+4 \, v_s^2 \, k^2}.
\end{eqnarray}

$\textbf{ii(b)}.~~$ However, for, ${\bf k}\, \parallel \, \bm B_0$ and $\delta \bm v_{\omega, {\bf k}}\, \perp \,\bm B_0$,
\begin{eqnarray}
\omega & = & - \frac{1}{2}\left( \frac{\xi_0 v_A}{(\epsilon_0+p_0)^{1/2}} k -i\frac{\eta_0 k^2}{(\varepsilon_0+p_0)}\right)\nonumber \\ 
\pm && \frac{1}{2}\sqrt{\left( \frac{\xi_0 v_A}{(\epsilon_0+p_0)^{1/2}} k -i\frac{\eta_0 k^2}{(\varepsilon_0+p_0)}\right)^2\, +\,4v_A^2 k^2}.
\end{eqnarray}
\subsection{Chiral Hydrodynamics with second order hydrodynamics of a chiral plasma}
In this case, we have considered the total energy momentum tensor:
\begin{equation}
T^{\mu\nu}=T_{\rm ideal}^{\mu\nu} + \tau_{(1)}^{\mu\nu} + \tau_{(2)}^{\mu\nu}
\end{equation}
where  $T_{\rm ideal}^{\mu\nu}$ and $\tau_{(1)}^{\mu\nu}$and $\tau_{(2)}^{\mu\nu}$ are defined in equations (\ref{T-munu}), (\ref{eq:Istorder}) and (\ref{eq:IIndorder}) respectively. To simplify the second order linearisation of the chiral hydrodynamics, in the present subsection,we have considered a homogeneous plasma for which $\nabla \mu =0$ in absence of electric field. The zeroth order and perturbed energy momentum tensor is
\begin{align}
T^{\mu\nu}_0 &=(\varepsilon_0+p_0) \, u_0^\mu u_0^\nu + p_0 \, g^{\mu\nu}\\
\delta T^{\mu\nu} & = (\varepsilon_0+p_0)\, [u_0^\mu \delta u^\nu+\delta u^\mu u_0^\nu] \nonumber\\
&-\eta_0 P^{\mu\alpha}_{0} P^{\nu\beta}_{0} [\nabla_\alpha\delta u_\beta+\nabla_\beta 
\delta u_\alpha]\nonumber\\
& +\frac{1}{3}\eta_0 P^{\mu\nu}_{0}\nabla_\alpha \delta u^\alpha 
+\lambda_{1_{0}} \Pi^{\mu\nu}_{{\alpha\beta}_{0}} \nabla^{\alpha}\delta\omega^\beta \nonumber\\
&+\lambda_{4_{0}}\Pi^{\mu\nu}_{{\alpha\beta}_{0}} \nabla^\alpha \delta B^\beta \label{eq:30}
\end{align}
The (00) and (0i) component of the perturbed energy momentum tensor can be found using Eq. (\ref{eq:30}) as
\begin{gather}
\delta T^{00}  = \delta \varepsilon \\
\delta T^{0i} =   (\varepsilon_0+p_0) \delta v^i
-\frac{\lambda_{{1}_{0}}}{2}\epsilon^{ijk}\partial_0\partial _j \delta v_k -
\frac{\lambda_{{4}_{0}}}{2}\partial_0 \delta B^i \\
\delta T^{ij} =  
 -\eta_0\large[g^{ik}\partial_k \delta v^j+g^{jk}\partial_k \delta v^i\large] +\frac{1}{3}\eta_0 g^{ij} \partial_k \delta v^k \nonumber \\
  +  \frac{\lambda_{{1}_{0}}}{2} [g^{ik}\epsilon^{jlm}\partial_k \partial _l \delta v_m + g^{jk} \epsilon^{ilm}\partial_k \partial _l \delta v_m -\frac{2}{3} g^{ij} \epsilon^{klm} \partial_k \partial_l \delta v_m] \nonumber \\
 +\frac{\lambda_{{4}_{0}}}{2}\large[g^{ik}\partial_k \delta B^j+g^{jk}\partial_k \delta B^i-\frac{2}{3}g^{ij}\partial_k\delta B^k\large]
\end{gather}
For a incompressible fluid ($\nabla\cdot \delta {\bf v}=0$), the $\nu =0$ component of the equation (\ref{conservT}) gives $\partial \delta \varepsilon/\partial \tau =0$.
Similarly for $\nu=j$, the conservation equation (\ref{conservT}) gives-
\begin{align}
(\varepsilon_0 & + p_0)\, \frac{\partial\, \delta\bm v}{\partial \tau} +\bm B_0\times\big(\bm\nabla\times\delta\bm B-\xi_0\,\delta \bm{\omega}\big) \nonumber \\
& = -\eta_0\, \nabla^2 \delta {\bf v}-\frac{\lambda_{1_0}}{2}\frac{\partial^2 \delta \boldsymbol{\omega}}{\partial \tau^2} - \frac{\lambda_{4_0}}{2}\frac{\partial^2 \delta {\bm B}}{\partial \tau^2} \nonumber \\
&+\frac{\lambda_{1_0}}{2}\nabla^2 \delta \boldsymbol{\omega} + \frac{\lambda_{4_0}}{2}\nabla^2 \delta {\bm B}. \label{secondordereuler}
\end{align}
Taking time derivative of equation (\ref{secondordereuler}) and replacing $\frac{\partial \delta {\bm B}}{\partial \tau}$ from equation (\ref{Maxwpert}) and taking Fourier transform
\begin{gather}
[-\omega^2 + \frac{i \eta_0 k^2}{\varepsilon_0+ p_0}\omega +\frac{\xi_0 (\bm v_A\cdot {\bf k})}{(\varepsilon_0 + p_0)^{1/2}} \omega- \frac{i \lambda_{{4}_{0}}(\bm v_A\cdot {\bf k})}{2(\varepsilon_0 + p_0)^{1/2}} \nonumber \\
 \times (\omega^2-k^2)]\delta \bm v
  =  \frac{\xi_0 (\bm v_A\cdot \delta \bm v) \omega}{(\varepsilon_0+ p_0)}\, {\bf k} +\frac{i\lambda_{{1}_{0}}}{2(\varepsilon_0 +p_0)}\,\nonumber \\
   \times(\omega^2 +k^2)\, \omega\, ({\bf k}\times \delta \bm v).
\end{gather}
Dot product of the above equation with ${\bf k}$ will give $\bm v_A \cdot \delta \bm v=0$. Now two cases are possible: ${\bf k} \parallel \delta \bm v$ and ${\bf k} \nparallel \delta \bm v$. \\

\textbf{Case~i.} When ${\bf k} \parallel \delta \bm v$. In this case, if ${\bf k}\cdot \bm v_A=0$, dispersion relation will be similar to the viscous incompressible viscous fluid, i.e. $\omega =\frac{i\eta_0 k^2}{\epsilon_0 +p_0}$. However, in the case of $\, ({\bf k}\cdot \bm v_A) \neq 0$,
\begin{align}
\omega & \approx \frac{i\eta_0 k^2}{2(\varepsilon_0+p_0)}+\frac{\xi_0 (\bm v_A \cdot {\bf k})k }{2(\varepsilon_0+p_0)^{1/2}} + \frac{\eta_0\,\lambda_{{4}_{0}} k^2  (\bm v_A \cdot {\bf k})}{4(\varepsilon_0+p_0)^{3/2}} \nonumber \\
&+ \frac{i \, \xi_0 \, \lambda_{{4}_{0}}\, k(\bm v_A \cdot {\bf k})^2}{2(\varepsilon_0+p_0)} + \mathcal{O}(k^4) 
\label{eq:secorder-kvneq0}
\end{align}
\textbf{Case~ii.} When ${\bf k} \nparallel \delta \bm v$, and $(\bm v_A \cdot {\bf k})= k_z B_0$,
\begin{eqnarray}
& &\frac{i \lambda_{{1}_0} k_z}{2 (\varepsilon_0+ p_0)^{1/2}}\omega^3  -  \omega^2\left(1+\frac{i \lambda_{{4}_0} v_A k_z}{2 (\varepsilon_0+ p_0)^{1/2}}\right)- \nonumber \\\nonumber
&  &\frac{i k_z}{2 (\varepsilon_0+ p_0)^{1/2}} (\lambda_{{1}_0}k^2-2\xi_0 v_A)\omega +\frac{i\lambda_{{4}_0}v_A k_z k^2}{2 (\varepsilon_0+ p_0)^{1/2}}=0\\
\end{eqnarray}
\section{A novel mechanism for Pulsar kicks}\label{sec-pulsarkick}
 We use our results for a qualitative calculation of the observed pulsar kick \cite{Minkowski:1970rm, Lyne:1982ab, Hansen:1997be}.  There are several attempts to explain the reason for the kick, for eg. see the references \cite{Gott:1970jr, Iben:1996ja, Kusenko:1999as,  Kusenko:2004mm}. Recently there have been attempts to explain the pulsar kick using chiral magnetic effect Ref. \cite{Kaminski:2014jda}. But the exact reason  for the pulsar kick is  still unknown.
 
 In order to calculate the pulsar kick, we consider the energy flux associated with the Alfv\'en mode. Further we note that the energy flux associated with the wave is energy density times the group velocity \cite{Stix:1992}, in our case which is the Poynting vector $\bm P=\bm E\times\bm B$ \cite{Stix:1992, Neil:2005} . In the momentum space, which can be expressed as,
 \begin{equation} 
 \bm P=(\omega  A^2)\bm k
 \label{Poynting_defi}
 \end{equation} 
 where $A$ is the magnitude of the vector potential $\bm A_{\omega,\bm k}$. Using (\ref{chiralalfvsplitmode}) we get \cite{Bhatt:2016irk},
 \begin{equation}
 \big| \bm P\big|=k^2A^2 v_g
 \label{poyntinggrp}
 \end{equation} 
 From Eqs. (\ref{chiralalfvsplitmode}) and (\ref{poyntinggrp}), we infer that there is a preferential transport of energy due to the difference in group velocity of the Alfv\'en modes. The additional amount of energy transported is given by,
 \begin{equation}
 \big|\Delta P\big|=\Big(\frac{\xi}{(\varepsilon_0+p_0)^{1/2}}\Big)(k^2A^2v_A) \label{eq:puls-kick-xi}
 \end{equation}
 This is essentially the excess amount of momentum carried by the photons leaving pulsar per unit area per unit time. Therefore the kick experienced by the pulsar can be written as,
 \begin{equation}
 \Delta V_{NS}=\frac{\Delta P}{M_{NS}}\times\Delta t\times \text{Area}
 \end{equation}
 where, $M_{NS}\sim 10^{30}$ kg is  mass of the neutron star, $\Delta t$ is the duration which  pulsar kick lasts and $\text{Area}=\pi r_N^2$, $r_N\sim 10$ km being the radius. For the order of magnitude calculation, we take $\xi\sim {1}/{2}C\mu_l^2$ where, $C={1}/{4\pi^2}$ is the anomaly coefficient and $\mu_l={\varepsilon}/{n_l}\sim 300$ MeV. We also assume the temperature $T\sim 10$ MeV and $k=A\sim T$. Using these values we get $\Delta V_{NS}\sim (10^2-10^3)$ km/s for $B_0\sim(10^{12}-10^{13})$ Gauss, which is within the limits of observed pulsar kicks. 
\section{Results and Discussions}

In this work, we have considered chiral hydrodynamics which may be relevant in treating plasma whose constituents particles  are massless. We have done the linear analysis to study the behaviour of the  modes allowed within such plasmas. We have shown  within the ideal hydrodynamic limit that, the longitudinal modes behaves in the same way as in standard plasma and the propagates with $|v_g| = \sqrt{v_s^2 +v_A^2}$, however,  the transverse mode, travel with different group velocities in the direction parallel and opposite to the background magnetic field. Since the group velocity depends on the anomaly coefficient $\xi \propto [\mu_R(T)-\mu_L(T))]$, this distinction vanishes only if  the plasma has no chiral imbalances. Using this result,  we have estimated  pulsar kick and shown that  the kick velocity $\Delta V_{NS} \text{ can be of the order of } 10^2-10^3$ km/s for a background field strength $B_0\sim (10^{12}-10^{13})$ Gauss (for strength of the background magnetic fields see \cite{Pandey:2017zpg}), which matches with the observations. Next we have considered chiral hydrodynamics with first and second order dissipative terms allowed by the conformal symmetry and done the linear analysis. We have shown that the collective modes of the chiral plasma get substantially modified in presence of the external magnetic fields due to the dissipative terms. For ${\bf k}\parallel {\bm B}_0$ and $\delta {\bm v}_{\omega, {\bf k}}\parallel {\bm B}_0$, in absence of the first order viscous term, we have sound waves propagating parallel  and anti-parallel to the external magnetic field. However, in presence of dissipative terms, the modes  damp due to  viscosity.  Furthermore, when ${\bf k}\parallel {\bm B}_0$ and $\delta {\bm v}_{\omega, {\bf k}}\perp {\bm B}_0$ the frequency $\omega$ get modifications from first order dissipative terms which in turn result in the  damping of the  modes.  In the case of ${\bf k} \perp {\bm B}_0$, sound waves and  Alf\'ven modes of the plasma get mixed up. However, they get damped at smaller scale due the the first order viscous corrections. It is to be noted here that, the chiral effects are seen only in the case of ${\bf k} \parallel {\bm B}_0$. The analysis done in the case of the second order hydrodynamics is quite completed than ideal chiral and first order hydrodynamics. We have considered that chiral potential and the temperature of the plasma is homogeneous (i. e. $\nabla \mu=0$ $\nabla T = 0$) and $\nabla\cdot \delta {\bm v}=0$. When ${\bf k}\cdot {\bm v}_A =0$ for ${\bf k}\parallel \delta {\bf v}$, modes of the plasma will only damps due to the viscosity. However, in the case of ${\bf k}\cdot {\bm v}_A \neq 0$, Alf\'ven modes modified due to the second order viscosity term. In equation (\ref{eq:secorder-kvneq0}), second term represent the chiral Alf\'ven term. Due to the second and third term, collective modes of the plasma grow. However, due to first and fourth term, these modes will be damped exponentially. This is one of the important result of our work. In the last part of the work, we have found a completed cubic dispersion equation.

So, to conclude the work, we have shown using first and second order conformal hydrodynamics that our results are consistent with the previous results. In the case of second order conformal hydrodynamics, for the chiral plasma, we have shown that the transport coefficients related with the second order conformal hydrodynamics actually contributes into the dispersion relations. We have calculated pulsar kick and we have found that the pulsar kick can be explained by chiral effects of the plasma at very high temperature. 
\section*{Acknowledgement} We would like to thank Dr. Jitesh R. Bhatt for carefully reading and for his fruitful discussions and suggestions. 
\bibliographystyle{apsrev4-1}{}
\bibliography{CH-MHD-pulsar.bib}
\end{document}